\documentclass[preprint,5p,times]{elsarticle}
\makeatletter
\def\ps@pprintTitle{%
	\let\@oddhead\@empty
	\let\@evenhead\@empty
	\def\@oddfoot{\reset@font\hfil\thepage\hfil}%
	\let\@evenfoot\@oddfoot
}
\makeatother

\usepackage{amssymb}

\usepackage{float}
\usepackage{amsmath}
\usepackage{algorithm}
\usepackage{algorithmic}

\usepackage{graphicx}
\usepackage{caption,subcaption, multirow}
\usepackage{makecell} 
\usepackage{array}    
\usepackage{colortbl} 
\usepackage{xcolor}   
\usepackage{booktabs}
\usepackage{adjustbox}  
\usepackage{multirow}
\usepackage{threeparttable}
\usepackage{hyperref}

\newcommand{\tmarkscale}{1.5}
\newcommand{\tmark}[2]{\kern0.1em\hyperlink{#1}{\smash[b]{\raisebox{0.1ex}{\scalebox{\tmarkscale}{\textbf{#2}}}}}}
\newcommand{\tanchor}[1]{\hypertarget{#1}{}}

\graphicspath{{./} }

\begin{document}

\begin{frontmatter}


\author{\texorpdfstring{Sudad Abed\corref{cor1}\fnref{aff1,aff2}}{Sudad Abed}}
\ead{s.abed@latrobe.edu.au}
\author[aff1]{Abdun Mahmood}
\author[aff1]{Mohammad Jabed Morshed Chowdhury}

\cortext[cor1]{Corresponding author.}
\affiliation[aff1]{organization={Department of Computer Science and Information Technology - La Trobe University},
            addressline={Moat Dr}, 
            city={Bundoora},
            postcode={3083}, 
            state={Victoria},
            country={Australia}}
\affiliation[aff2]{organization={Electronic Computer Center - University of Anbar},
    addressline={55 St.}, 
    city={Ramadi},
    postcode={31001}, 
    state={Anbar},
    country={Iraq}}

\title{PoCQ: Defending Decentralised Federated Learning Against Model Poisoning and Collusion with Verifiable Evidence}
\tnotetext[notes]{This work has been submitted for possible publication.}

\begin{abstract}
Decentralised federated learning removes the central coordinator but requires participants to establish model update integrity autonomously. Existing frameworks either use inexpensive yet unverifiable proxies, including dataset size and epoch count, or validate updates through retraining, which is computationally costly. This paper introduces Proof of Contribution Quality (PoCQ), a framework in which trust is based on published evidence rather than on private judgement. Updates are committed before disclosure and assessed by small assigned peer committees against the updates each committee already holds. No common validation dataset is required, no validator retrains the evaluated model, and no participant requires a view of all updates in a round. Every vote is published with the values that produced it, so any peer can recompute it and contradict a dishonest validator. Accountability is graded by the strength of that evidence, with permanent removal reserved for provable misconduct and bounded suspensions for statistical anomalies, which honest nodes can produce under non-IID data. Across five model poisoning attacks on PathMNIST, PoCQ improves the average update level detection precision of the best current framework by 49\% under non-IID data and by 206\% under IID, and improves the precision of malicious node detection by 106\% under non-IID data. This is achieved while reducing the time required to validate by half against the fastest existing framework and by 92\% against retraining based validation. Colluding bad-mouthing validators are removed by recomputing their published evidence, and free riders who replay or copy an update exactly are identified directly from the ledger.
\end{abstract}

\begin{keyword}
decentralised Federated Learning \sep Trustworthy AI\sep Reputation Systems\sep Model Poisoning\sep Verifiable Computation.
\end{keyword}

\end{frontmatter}

\section{Introduction}
The digital transformation of healthcare has created vast repositories of sensitive patient data, ranging from medical imaging to electronic health records (EHRs). While these datasets hold immense potential for training diagnostic artificial intelligence models, firm privacy regulations, including HIPAA and the GDPR, create ``data silos,'' preventing institutions from pooling their information centrally \cite{Joshi2022}. Federated Learning (FL) has been introduced as a critical solution to this dilemma, it enables hospitals and medical centres to collaboratively train global models by exchanging only parameter updates, while patient data remains securely within the institution's firewall \cite{McMahan2017}. However, the traditional FL architecture depends completely  on a central aggregation server. In a healthcare context, this centralised design presents a high-risk Single Point of Failure (SPoF) and a clear aim for cyberattacks. To mitigate the problem of SPoF, recent research has pivoted toward decentralised Federated Learning (DFL) and optimised methods to speed the learning process \cite{liu2022decentralised, abed2025t}. In addition, recent research has used blockchain technology to replace the central authority with a peer-to-peer consensus network to improve resilience and trust \cite{kim2019blockchained, li2020blockchain}. While decentralisation mitigates the single point of failure issue, it introduces severe adversarial dynamics in a trustless environment. Malicious participants may launch model poisoning attacks, injecting corrupted gradients to weaken the performance of the final model \cite{xia2023poisoning}. In addition, the consensus mechanism itself is vulnerable to bad-mouthing attacks, where dishonest validators deliberately reject valid updates from rival clients to damage their reputation \cite{lewis2023attacks}. Furthermore, within a competitive healthcare environment, certain institutions may perform free-rider attacks by appearing to participate in the federated learning process while ultimately aiming to acquire the aggregated global model without computational effort \cite{fraboni2021free}.
	
Existing Blockchain-based FL (BCFL) frameworks have attempted to secure this infrastructure, though often with four limitations that affect their viability in clinical settings. The first limitation is that a validation decision cannot be recomputed by the peers who have to accept it, and that it rests either on quantities the worker reports about itself or on a test set only the validator can see. For instance, LBFL \cite{qiao2024lbfl} employs a committee-based consensus to improve throughput and storage efficiency. In addition, LBFL reputation score calculations depends not only on the improvement of global model based on the updates of a client but also on the number of samples that the client owns and the number of epochs performed, which is hard to validate. Conversely, VBFL \cite{chen2021robust} prioritises security by introducing a validator role to check the quality of received updates before aggregation. However, VBFL relies on an accuracy-based validation mechanism, requiring validators to re-train received updates on their local validation datasets, so every validator must hold a test set and must retrain the model it is judging. Both designs also expect a validator to see the whole round of updates, so they remove the central server without removing the central view.
	
The second limitation concerns the quality of the judgement that follows. Beyond deliberate attacks, real world medical data is rarely uniform or Independent and Identically Distributed (IID) \cite{lu2024federated}. This uneven data distribution causes model drift and reduces overall accuracy. A defence that cannot separate this skew from poisoning therefore penalises honest institutions for the shape of their data, which is the precision gap that existing frameworks leave open.
	
The third limitation is the cost of reaching that judgement. The re-training process imposes a significant computational overhead and latency, which can bottleneck the training of models on large-scale medical images. The fourth limitation is that the consensus layer itself is left unguarded. While LBFL effectively addresses the scalability of storing model updates, its primary focus is on system performance rather than cryptographic defence against different attacks or the sophisticated manipulation of reputation scores by colluding validators, and neither framework identifies a participant that replays or copies an update from the ledger instead of training it.
	
A separate body of work defends the aggregation step itself. Krum, coordinate wise trimmed mean and coordinate wise median replace the average by a statistic that a minority of corrupted inputs cannot move far \cite{blanchard2017krum, yin2018byzantine}. These rules are aggregation rules and not detection mechanisms, since they suppress the influence of an anomalous update within a round without forming any persistent judgement about the node that produced it. They also assume that one party holds every update of the round, which is precisely the assumption that decentralisation removes. They therefore serve as a reference point for robustness in this work rather than as competing malicious detection designs.
	
A third group judges an update by testing it on a dataset the validator already holds. FLTrust gives the server a small clean root dataset and scores each update by how closely it agrees with a reference model trained on that data \cite{cao2021fltrust}. T-BFL needs no server and instead gives every node the same public dataset, so each node can test every peer and any node can reproduce the result \cite{Zhu2025}. Both therefore rely on labelled data that the validators hold before training starts. DeFL takes a different route, detecting poisoning at the server from the layer norms of every client update rather than from a dataset, and it notes that a server often has no such dataset to begin with \cite{Yan2023}, but it still requires one party to see the whole round. In a hospital federation such a dataset is harder still to obtain, since the records that would form it cannot leave the institutions holding them. PoCQ is built so that neither the dataset nor the whole round view is needed, and these methods are therefore not evaluated here.
	
To address these four limitations in turn, we propose Proof of Contribution Quality (PoCQ), a framework for trustworthy decentralised federated learning in which every validation decision is published as evidence that peers can recompute or challenge. A PoCQ validator judges a peer against the small set of updates it already holds, using a per layer norm profile that detects a change in magnitude together with a directional test that detects a change in orientation. These two trust metrics remove the three requirements that limit existing validation. No common validation dataset has to be assembled and held by the participants, so the privacy preserving goal of federated learning is not defeated. No validator retrains the evaluated model, so the overhead of validation stays low. No participant requires a view of all updates in a round, which existing designs still require after removing the central server. The quality of that judgement is preserved because penalties depend on how strong the evidence is. A node is permanently removed only when the ledger itself clearly shows that it violated the rules, while an institution whose data is merely skewed receives a bounded suspension instead. The cost of reaching that judgement is four layer norms and one cosine per update rather than a training epoch, unlike the computationally intensive re-training of VBFL. The consensus layer is guarded as well. Unlike LBFL, PoCQ integrates a strict cryptographic protocol to prevent manipulation of the reputation score, so a colluding validator is exposed by recomputing its own published evidence and a replayed or copied update is recognised from the ledger, ensuring that credit for clinical contributions is correctly attributed. The evaluation then follows the same order, establishing the detection quality of this rule, the cost of applying it, and its behaviour under attacks directed at the consensus layer itself. 
	
PoCQ is evaluated against five model poisoning attacks and two consensus layer attacks, detailed in Section~\ref{sec:attacks}. Gaussian noise and sign flipping are blind and complementary, one inflating magnitude and the other reversing direction while preserving every norm, so no single statistic detects both. LIE \cite{baruch2019lie}, Min-Max \cite{shejwalkar2021manipulating} and Fang \cite{fang2020local} are informed, crafted to sit inside the range a defence accepts. Bad-mouthing \cite{lewis2023attacks} and free-riding \cite{fraboni2021free} target the protocol rather than the model. Together they cover the three ways a decentralised framework fails, a corrupted model, a corrupted vote and an absent contribution.
	
The specific contributions of this paper can be highlighted as follows:

\begin{itemize}

\item Introduces PoCQ, a decentralised federated learning framework in which each peer validates updates from a subset of peers. Every vote is published with its evidence, a layer norm profile for magnitude and a cosine similarity for direction, so any peer can recompute it. This removes three requirements of existing validation approaches, a common validation dataset, which defeats the privacy preserving goal of federated learning; retraining of the evaluated model, which adds overhead; and a full view of the round, which limits scalability.
		
\item Improves the average update-level detection precision across the five attacks against the best current frameworks (VBFL and LBFL) for non-IID data by 49\% and for IID data by 206\% (Table~\ref{tab:detection}). The approach also improves the precision of malicious node detection by 106\% for non-IID data (Table~\ref{tab:node_precision} and Fig.~\ref{fig:confusion_m}).		
\item Improves validation efficiency by reducing time required to validate by half compared to the fastest technique (LBFL) and by 92\%  compared to the retraining-based validation technique (VBFL) (Fig.~\ref{fig:val_time}).
		
\item Addresses the issue of collusion among bad-mouthing validators, which affects existing approaches, by recomputing their published evidence using the trust metrics mentioned above. It is also capable of identifying free rider who replay and copy updates directly from the published ledger (Table~\ref{tab:consensus_attacks}).
\end{itemize}

The remainder of this paper is structured into four primary areas. Section \ref{sec:related_work} reviews the current advancements in secure federated learning. Section \ref{sec:methodology} comprehensively details the PoCQ framework and its five phases and the underlying system architecture. Section \ref{sec:experiments} presents the experimental setup alongside a rigorous analysis of the performance results. Finally, Section \ref{sec:conclusion_future_work} summarises the core findings of the study and outlines promising directions for future research.

\section{Background and Related Work}
\label{sec:related_work}
This section presents federated learning from centralised to blockchain enabled decentralised architectures, the attacks that decentralisation exposes, and the limits of existing consensus, aggregation and reputation defences.

\subsection{decentralised Federated Learning in Healthcare}
Federated Learning keeps patient records inside the institution, but its conventional client-server form concentrates aggregation in a single party, which introduces a single point of failure, a scalability bottleneck and exposure to server side attacks \cite{McMahan2017, nezhadsistani2025blockchain}. Yuan et al. \cite{yuan2024decentralised} show that decentralised Federated Learning removes that party by distributing the aggregation workload across peer nodes through gossip protocols or structured topologies. While this improves robustness against server failure, it fundamentally alters the trust model, necessitating new mechanisms to verify the integrity of contributions in the absence of an authoritative coordinator.

\subsection{Model Poisoning and Consensus Attacks}
\label{sec:attacks}

The transition to a trustless, decentralised environment exposes the learning process to severe adversarial threats. Recent studies classify these threats into two primary categories, performance degradation and reputation manipulation \cite{hallaji2024decentralised, beltran2023decentralised}. A decentralised setting adds a third concern, participation without contribution.
    
First, model poisoning attacks involve malicious clients injecting mathematically corrupted gradients to disrupt convergence of the global model or to implant targeted backdoors \cite{hallaji2024decentralised, Yan2023}. In high-dimensional spaces typical of medical imaging, such attacks can be subtle; adversaries may scale their updates to bypass simple magnitude checks while still misdirecting the optimisation trajectory. Conversely, in competitive cross institutional settings, the system faces free-rider attacks. As demonstrated by Fraboni et al. \cite{fraboni2021free}, selfish participants may download the global model and upload random or repeated weights to simulate participation. This behaviour drains network bandwidth and dilutes the global model quality without contributing legitimate data.

Model poisoning attacks can be classified according to the amount of information available to the adversary during a training round. In a blind attack, the adversary changes only its own update without considering the updates of other participants. Examples include adding Gaussian noise to the trained weights or reversing the sign of every parameter. they are inexpensive and require no coordination, but they are also the most exposed to a statistical test, because the resulting update carries no information about the characteristics of normal updates in the same round. LIE ``a little is enough'', Min-Max and Fang are the informed attacks evaluated in this work, and they are defined in Section~\ref{sec:threat_model} \cite{baruch2019lie, shejwalkar2021manipulating, fang2020local}.

The distinction matters for a decentralised protocol, because an informed attack presupposes a full view of the round that a committee based design does not grant to any participant.

A more insidious threat in decentralised consensus is the bad-mouthing attack \cite{lewis2023attacks}. As detailed by Zhang et al. \cite{zhang2025trust} in their taxonomy of trust attacks, malicious validators may collude to cast negative votes against valid updates submitted by honest nodes. The objective is to artificially degrade the reputation of rival institutions, effectively ejecting them from the consensus committee to monopolise mining rewards or influence. Unlike poisoning, which attacks the model, bad-mouthing attacks the consensus mechanism itself, requiring defences that can objectively verify validator honesty.

\subsection{Blockchain-Based Consensus Mechanisms}
To enforce trust and auditability in DFL, researchers have integrated blockchain technology, creating Blockchain-Based Federated Learning (BCFL). The immutable ledger provides a tamper-proof history of model updates and participant behaviour, yet existing frameworks exhibit distinct trade-offs between security and efficiency. Early versions, such as Blockchained On-Device FL \cite{kim2019blockchained}, relied on Proof-of-Work (PoW) consensus. While secure, PoW imposes prohibitive computational costs and latency, making it unsuitable for the iterative, high-frequency communication required by deep learning. Consequently, the field has shifted toward committee-based mechanisms to improve throughput, as seen in the work of Li et al. \cite{li2020blockchain}, where a randomly selected subset of nodes validates updates. However, random selection does not guarantee the honesty of the committee, leaving the system vulnerable to collusive attacks.

To explicitly address the quality of updates, Chen et al. \cite{chen2021robust} proposed VBFL, which mandates that validators re-train received updates on their local validation datasets to verify accuracy improvement. While this ``Proof-of-Retraining'' detects attacks that degrade accuracy on the validator's own data, it incurs a massive computational overhead; the validation time scales linearly with model complexity, creating a severe bottleneck for large medical models where re-training a single update can take minutes. Addressing this efficiency gap, Qiao et al. \cite{qiao2024lbfl} introduced LBFL, a lightweight framework employing a Proof-of-Contribution (PoC) consensus that defines contribution not only by retraining the model for one epoch by validators but also based on data volume and number of training epochs by a worker node. LBFL uses one miner only in each iteration, which receives votes for trained workers' models from a selected validation committee to reduce the communications. Similarly, Zhao et al. \cite{zhao2024long} introduced a ``Long-Term Proof-of-Contribution'' algorithm designed to incentivise sustained participation over time rather than immediate utility. Most recently, I\c{s}ler et al. \cite{icsler2025fedpop} introduced FedPoP, utilising cryptographic proofs to verify that local training physically occurred, thereby effectively preventing free-riding.
    
However, these efficiency focused frameworks exhibit a critical limitation: they largely define contribution quantitatively rather than qualitatively. Frameworks like LBFL and FedPoP lack rigorous mechanisms to verify the semantic quality of updates against intelligent poisoning. For instance, FedPoP ensures a client performed work, but cannot distinguish between work done on legitimate data versus corrupted data. Furthermore, LBFL relies on majority voting without specific defences against bad-mouthing attacks where validators manipulate consensus scores. In addition, the reputation score mechanism in LBFL relies on metrics that are inherently difficult to verify, such as the client's local dataset size and the number of training epochs performed.

A third design stays decentralised but gives the validators data to judge with. T-BFL removes the central server and evaluates reputation by having every node test each peer's update on the same public dataset, so any node can reproduce another node's score \cite{Zhu2025}. That reproducibility is the property PoCQ also seeks, but T-BFL obtains it by assuming a public dataset is already present on every node, which a hospital federation cannot assume. PoCQ derives its evidence from the updates themselves, so a score stays reproducible without that dataset.
    
Both VBFL and LBFL have a structural limitation. Each validator bases its vote on accuracy measured using private validation data. Although the vote is recorded on the ledger, other nodes cannot verify the underlying evidence. As a result, dishonest voting is managed only through reputation, which a sufficiently large number of malicious nodes can preserve. Our approach publishes the values used to produce each vote, allowing every node to independently recompute most of the vote without relying on the validator.

\subsection{Robust Aggregation Rules}
\label{sec:robust_agg}
Robust aggregation provides a separate defence by reducing the influence of poisoned updates at the aggregation stage. Krum selects the update closest to its nearest neighbours \cite{blanchard2017krum}. Coordinate-wise trimmed mean and median remove or ignore extreme values in each parameter dimension \cite{yin2018byzantine}. FLTrust weights updates according to their similarity to a reference update generated by the server using a small, clean root dataset \cite{cao2021fltrust}.

Unlike PoCQ, these methods are aggregation rules rather than detection mechanisms. They may down-weight anomalous updates, but they do not assess node behaviour across rounds, build reputation, or exclude participants. They also require one party to access all updates in a round. Krum and trimmed mean additionally require the number of adversaries, which conflicts with a peer-to-peer setting.

We therefore use Krum, trimmed mean, and coordinate median as accuracy references in Section \ref{sec:experiments}. We provide them with a full view of the population and the true number of adversaries. Their detection metrics are undefined because they do not detect malicious nodes.

\subsection{Reputation Systems and the Non-IID Challenge}
To move beyond simple voting, reputation-aware systems have been developed to penalise poor performance and ensure fairness \cite{Zhu2025, barkatsa2025fair}. However, a profound limitation of current decentralised reputation frameworks is their inability to efficiently manage statistically heterogeneous, or non-IID, data.

Real-world medical data is rarely uniform \cite{lu2024federated}. Training on non-IID data inherently causes model drift and uneven local updates, which traditional reputation metrics frequently misclassify as malicious poisoning. Recent advancements have attempted to distinguish between adversarial behaviour and natural statistical divergence. For example, Kasyap et al. \cite{kasyap2023efficient} proposed the Proof of Interpretation and Selection (PoIS) consensus, utilising model interpretation techniques (Shapley values) to fairly evaluate a client's contribution across specific data distributions.
    
While interpretation based mechanisms like PoIS provide deep, data-aware security, they introduce a prohibitive computational burden. The calculation of feature attributions scales exceptionally poorly with high-dimensional deep neural networks, rendering it practically infeasible for the rapid, lightweight validation required in clinical edge environments.

A second difficulty is rarely separated from the first. A reputation system that removes a node permanently on statistical evidence treats an inference about behaviour as though it were a proof, and under non-IID data the inference is sometimes wrong. The cost of that error is not symmetric, because an honest node removed from the network takes its data with it and the global model never recovers the contribution. This observation motivates the two tier design described in Section \ref{sec:methodology}, in which permanence is reserved for evidence that no honest node can produce.
    
Consequently, the current literature lacks a consensus mechanism capable of providing the semantic security of robust frameworks while maintaining the low-latency throughput of lightweight models. To resolve this fundamental limitation, we propose PoCQ, a reputation-aware consensus mechanism that leverages lightweight geometric auditing to objectively verify update quality. By bypassing the need for heavy re-training or deep model interpretation, PoCQ simultaneously neutralises poisoning, mitigates bad-mouthing, and maintains robust performance in non-IID healthcare environments.

\section{The PoCQ Framework}
\label{sec:methodology}

Removing the central server also removes the party that decided which updates entered the global model. A decentralised protocol must therefore answer three questions on its own. Was an update formed honestly, can the judgement passed on it be trusted, and what should happen to a participant that fails either test. PoCQ answers all three from evidence the ledger carries, rather than from the private opinion of whichever peer happened to look.

Three design choices follow, each addressing a limitation discussed in Section~\ref{sec:related_work}. First, an update is committed before it is revealed and is then judged against the small set of updates its validator already holds, using a geometric test that needs one pass over the parameter vector and no validation data. Second, every vote is published together with the values that produced it, so the magnitude component of any vote can be recomputed by any node and a dishonest validator is contradicted by its own record. Third, penalties are graded by the strength of the evidence, so a violation provable from the ledger leads to permanent removal, while a statistical anomaly, which an honest node can also produce under non-IID data, reduces reputation and may lead to a bounded suspension.
    
\begin{figure*}[ht]
    \centering
    \includegraphics[scale=0.8]{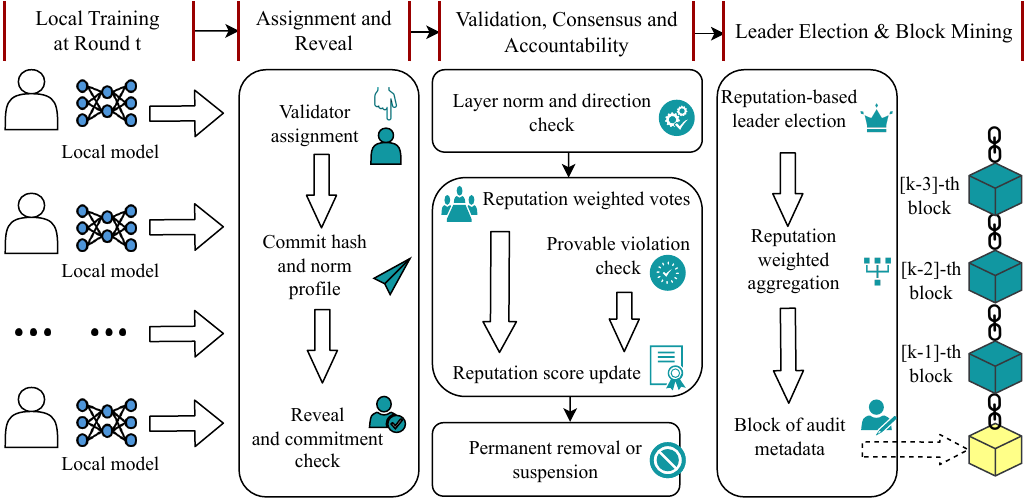}
    \caption{Workflow of PoCQ.}
    \label{fig:Methodology}
\end{figure*}

PoCQ runs over a peer-to-peer network secured by a blockchain. Identity rests on a Public Key Infrastructure, in which every node $i$ holds a key pair $\{PK_i, SK_i\}$ and the public key serves as the node identity. Every communication round $t$ proceeds in five synchronized phases, shown in Fig.~\ref{fig:Methodology}, which together implement these three answers. Table~\ref{tab:notations} lists the notation used.

\begin{table}[ht!]
\centering
\caption{Summary of Key Notations}
\label{tab:notations}
\small
\begin{tabular}{@{} l p{0.78\columnwidth} @{}}
\toprule
\textbf{Notation} & \textbf{Description} \\
\midrule
$t$, $T_{\max}$ & Communication round index and total number of rounds \\
$i, j$ & Node indices, worker $i$ and validator $j$ \\
$PK_i$, $SK_i$ & Public and private key of node $i$ \\
$w_g^t$ & Global model at round $t$ \\
$\Delta w_i^t$ & Local model update of node $i$ at round $t$ \\
$D_i$ & Local dataset of node $i$ \\
$H_i$ & SHA-256 hash of $\Delta w_i^t$, published as the commitment \\
$sig_i$ & Digital signature of node $i$ \\
$\Lambda$, $\ell$ & Number of weight matrices in the model and their index \\
$\mathbf{p}_i$ & Norm profile of $\Delta w_i^t$, one entry per weight matrix\\
$\mathcal{V}_i$ & Set of validators assigned to worker $i$ \\
$k_{\min}$, $k_{\max}$ & Lower and upper bounds on the size of $\mathcal{V}_i$ \\
$\mathcal{H}_j$ & Set of updates held by validator $j$ in the current round \\
$\mathcal{R}_{j \to i}$ & Reference used by $j$ to judge $i$, obtained by leaving $i$ out of $\mathcal{H}_j$ \\
$\mathbf{m}_{j \to i}$ & Median profile of $\mathcal{R}_{j \to i}$ \\
$\mathbf{d}_{j \to i}$ & Median direction of $\mathcal{R}_{j \to i}$ \\
$r_{\ell}$ & Norm ratio at weight matrix $\ell$ \\
$\tau$, $\zeta$ & Upper and lower bounds of the accepted ratio range \\
$\nu$ & Number of flagged weight matrices required to reject \\
$\phi$, $\phi_{\min}$ & Cosine similarity to the reference direction and its lower bound \\
$z_{j \to i}$ & Vote cast by validator $j$ for worker $i$, $1$ valid and $0$ invalid \\
$\mathcal{Z}_{tx}$ & Signed vote transaction recorded on the ledger \\
$S_i$ & Reputation weighted consensus score for worker $i$ \\
$\Psi_i$ & Consensus decision for worker $i$, $1$ accepted and $0$ rejected \\
$R_i^t$ & Reputation score of node $i$ at round $t$, $R_i^t \in [0,1]$ \\
$\beta$ & Reputation decay factor \\
$T_{\mathrm{warm}}$ & Warm-up rounds before reputation is scored \\
$R_{\min}$ & Reputation threshold that triggers suspension \\
$T_{\mathrm{susp}}$ & Suspension length, doubling on each repeat \\
$P_{\mathrm{elect}}(i)$ & Probability of node $i$ being elected round leader \\
$M$ & Number of adversarial nodes in the network \\
\bottomrule
\end{tabular}
\end{table}

\subsection{Phase 1. Local Training, Commitment and Reveal}
Worker $i$ trains the current global model on its own local dataset by Stochastic Gradient Descent and forms the update
\begin{equation}
\Delta w_i^t = w_i^{t+1} - w_g^t
\end{equation}

The update is described by a compact descriptor that the remaining stages of the protocol work with, rather than the update vector itself. A model is made of $\Lambda$ weight matrices, and the update contains one change per matrix. Taking the $L_2$ norm of each change gives one magnitude per matrix, and collecting them gives the norm profile

\begin{equation}
\mathbf{p}_i = \Big( \|\Delta w_{i,1}^t\|_2, \ \|\Delta w_{i,2}^t\|_2, \ \dots, \ \|\Delta w_{i,\Lambda}^t\|_2 \Big)
\end{equation}
where $\Delta w_{i,\ell}^t$ denotes the part of the update belonging to weight matrix $\ell$. Since every block holds many parameters and the norm returns one number, the profile summarises the update without reproducing it. For the network used in this work it stands for roughly $1.66$ million values in $\Lambda = 4$ magnitudes. The profile is a refinement of the familiar whole model norm rather than a different quantity, because

\begin{equation}
\|\Delta w_i^t\|_2 = \|\mathbf{p}_i\|_2
\end{equation}
Separating them allows the Phase 3 validation rule to determine how many parts of the model disagree, rather than merely measuring the overall magnitude of model change. Per layer inspection is also used by DeFL \cite{Yan2023}, although there it is computed at a central server holding every client update, whereas here each validator uses only the reference it already holds.

Each round proceeds through a commit step and a reveal step. During the commit step, worker $i$ publishes the hash of its update and its profile $\mathbf{p}_i$ in a message signed with its private key,
\begin{equation}
H_i = \text{Hash}(\Delta w_i^t), \qquad sig_i \leftarrow \mathsf{Sign}_{SK_i}\big( \{H_i, \mathbf{p}_i\} \big)
\end{equation}
The reveal step begins only after the commitment block has closed. The worker then sends the update to its assigned validators, which are already fixed at the start of the round by the previous block hash, as Phase 2 describes. Each validator recomputes the hash and profile and checks them against the published values $H_i$ and $\mathbf{p}_i$.

This ordering prevents adaptive attacks. Since all commitments are sealed before any update is disclosed, an adversary in round $t$ can rely only on information from round $t-1$ and cannot fit its submission to the honest updates of the current round. The commitment record also identifies two forms of free-riding directly from the ledger. Reusing a commitment previously submitted by the same node constitutes a replay, whereas submitting a commitment first published by another node constitutes a copy. Neither case requires any update to be examined.

Publishing $\mathbf{p}_i$ requires $\Lambda$ values per node per round and reveals no parameter values. It allows every node to reproduce the Phase 3 magnitude test, although updates are disclosed only to their assigned validators. Phase 4 relies on this property.

\subsection{Phase 2. Validator Assignment}
Each worker $i$ is assigned a validator set $\mathcal{V}_i$ whose size is drawn independently for that worker inside the bounds
\begin{equation}
    k_{\min} \le |\mathcal{V}_i| \le k_{\max}
\end{equation}
The assignment is a deterministic function of the previous block hash, so any node can reconstruct the entire mapping from the chain and no node can influence who audits it. Assignments are made in ascending order of current load, which spreads the work evenly, so that a validator judges approximately $|\mathcal{V}_i|$ peers per round.

Since the mapping is public and recomputable, a validator cannot deny having been assigned a peer, and any node knows exactly which updates a given validator held. Because a validator receives several updates in the course of its own duties, it holds a small sample of the round at no additional communication cost, and that sample is the reference of Phase 3.

\subsection{Phase 3. Reference Based Validation}
Let $\mathcal{H}_j$ be the set of updates that validator $j$ holds in the current round, namely those of the peers assigned to it together with its own. To judge worker $i$, the validator uses the leave-one-out reference
\begin{equation}
    \mathcal{R}_{j \to i} = \mathcal{H}_j \setminus \{i\}
\end{equation}
This ensures that the candidate update is excluded from the reference set. The composition of this set determines the size of the adversarial coalition that the mechanism can tolerate, as discussed further in Section \ref{sec:stress}.

Because $\mathcal{R}_{j \to i}$ is a set rather than a single peer, a summary statistic is available, and PoCQ uses the median, which is the location estimator that tolerates the largest fraction of corrupted samples. Validation is purely geometric on the parameter vectors. No validator retrains, and no validator evaluates a peer on local data. A norm or a cosine costs $\mathcal{O}(d)$ for a model of $d$ parameters, so the check completes in a fraction of one training step.

The following tests capture the two ways an update can differ from its peers, namely magnitude and direction. The norm profile captures magnitude, while the cosine captures direction, representing a $d$-parameter update with $\Lambda + 1$ values.

\subsubsection{Magnitude test}
Let $\mathbf{m}_{j \to i}$ be the entry wise median profile of the reference. The ratio at weight matrix $\ell$ is
\begin{equation}
    r_{\ell} = \frac{\mathbf{p}_{i,\ell}}{\mathbf{m}_{j \to i,\ell}}
\end{equation}

and a weight matrix is flagged when this ratio leaves a band with both an upper and a lower bound,

\begin{equation}
    \text{flag}(\ell) = \mathbb{I}\big( r_{\ell} > \tau \ \lor \ r_{\ell} < \zeta \big)
\end{equation}

The upper bound $\tau$ detects updates with abnormally large magnitudes, as commonly produced by noise injection or scaling attacks. The lower bound $\zeta$ identifies updates too small to result from genuine training. It therefore detects free-riders whose modified copy falls below that bound, although a perturbation that preserves the norm is not caught, as Section~\ref{sec:consensus_attacks} reports. An update is deemed invalid only when at least $\nu$ matrices are flagged.
    
\begin{equation}
    \text{layers\_ok} = \mathbb{I}\Big( \textstyle\sum_{\ell=1}^{\Lambda} \text{flag}(\ell) < \nu \Big)
\end{equation}

Requiring $\nu$ flagged matrices rather than one makes the test robust to heterogeneous data. An honest node with a different label distribution may typically deviate in only one matrix, whereas a poisoned update affects several parts of the profile at the same time.

Published Phase 1 profiles and Phase 2 assignments allow any node to recompute the median profile, the ratios, and the outcome without holding a single update.

\subsubsection{Direction test}
Since negating every parameter leaves all norms unchanged, magnitude alone cannot detect attacks that preserve the norm while reversing or rotating the update. PoCQ therefore applies a second, independent directional test. Let $\mathbf{d}_{j \to i}$ denote the coordinate wise median of the reference updates. The candidate must not point against it.
\begin{equation}
    \phi = \cos\big(\Delta w_i^t, \ \mathbf{d}_{j \to i}\big), \qquad \text{dir\_ok} = \mathbb{I}\big( \phi \ge \phi_{\min} \big)
\end{equation}

The threshold $\phi_{\min}$ is set slightly below zero. Thus, an update that is simply uncorrelated with its peers, as can occur under non-IID data, passes, whereas an update directed against the consensus direction does not. Unlike the magnitude test, this one is computed from the updates themselves. It can therefore be reproduced only by a node that holds them, and Phase 4 settles it by challenge rather than by public recomputation.

\subsubsection{Vote}
The validator accepts only if both tests pass,
\begin{equation}
    z_{j \to i} =
    \begin{cases}
    1 \ (\text{valid}) & \text{if layers\_ok} \ \land \ \text{dir\_ok} \\
    0 \ (\text{invalid}) & \text{otherwise}
    \end{cases}
\end{equation}
The vote is published together with the ratios and the cosine that produced it, in a signed and timestamped transaction,
\begin{equation}
    \mathcal{Z}_{tx} = \mathsf{Sign}_{SK_j}\big( \{ \text{ID}_{i}, \text{ID}_{j}, z_{j \to i}, \{r_{\ell}\}, \phi, \text{timestamp} \} \big)
\end{equation}

\subsection{Phase 4. Consensus and Accountability}
This phase first resolves provable violations, then aggregates the votes, and finally updates reputation. Since such violations are settled through recomputation rather than voting, they are resolved before aggregation, so no consensus decision can be influenced by dishonest validators.

\subsubsection{Tier one, provable violations}
A node is permanently removed when the ledger contradicts it. Five checks meet this condition, none of which can be triggered by an honest node. Removal follows when a revealed update does not hash to the published $H_i$, when it does not reproduce the published $\mathbf{p}_i$, when a commitment matches one previously submitted by the same node or one first submitted by another node, or when a published vote conflicts with the vote implied by the validator's own published ratios and cosine values. The last check prevents bad-mouthing. Because profiles and assignments are public, any node can recompute the magnitude component of a vote at any time, so a validator that votes invalid against an honest worker on magnitude grounds is immediately contradicted, without requiring cooperation from any other participant.

\subsubsection{Challenging a vote}
Because the direction test cannot be verified from public data alone, it is resolved only when challenged. A worker whose update is rejected may challenge the vote cast against it. The challenged validator must then disclose the inputs used for the test, including its own update and the updates of peers in its reference set. The ledger records the assignment, and commitments bind all updates, so the reference membership is known and its contents cannot be altered after the fact. Any node can then recompute the median direction and cosine values and compare them with the published values. A validator whose published values conflict with its inputs is subject to tier one removal, while failure to respond constitutes failure of the challenge.

Only the rejected worker may challenge a vote, and only the vote concerning its own update. The number of challenges in a round can therefore not exceed the number of rejections. A worker whose challenge fails receives a reputation penalty, which discourages challenges intended only to burden the network.

\subsubsection{Consensus decision}
Votes cast by nodes removed this round are discarded. The remainder are combined in proportion to validator reputation,
\begin{equation}
    S_i = \frac{\sum_{j \in \mathcal{V}_i} z_{j \to i} \cdot R_j}{\sum_{j \in \mathcal{V}_i} R_j}, \qquad
    \Psi_i =
    \begin{cases}
    1 \ (\text{accepted}) & \text{if } S_i > 0.5 \\
    0 \ (\text{rejected}) & \text{otherwise}
    \end{cases}
\end{equation}
Reputation weighting gradually reduces the influence of validators whose contributions are rejected. As adversarial nodes are detected, they lose voting power rather than retaining it.

\subsubsection{Tier two, statistical outcomes}
A rejected update is evidence and not proof, because an honest node holding non-IID local data can be rejected legitimately. Accordingly, this tier does not permanently remove a node. Each node receives one reputation step per round, determined by the consensus decision on its own update and applied as an exponential moving average with decay $\beta$ \cite{brotons2024exponential},

\begin{equation}
    R_i^{t+1} = \max\Big(0, \ \min\big(1, \ \beta R_i^{t} + (1-\beta)\,\rho_i \big)\Big), \qquad
    \rho_i =
    \begin{cases}
    +1 & \text{if } \Psi_i = 1 \\
    -1 & \text{if } \Psi_i = 0
    \end{cases}
\end{equation}

Reputation is frozen for the first $T_{\mathrm{warm}}$ rounds, since early updates are volatile under non-IID data and allowing rejections to drain reputation during that period would leave honest nodes at the floor by the time enforcement begins.

After the warm-up period, a node whose reputation falls below $R_{\min}$ is suspended rather than permanently removed. It cannot submit updates or cast votes for $T_{\mathrm{susp}}$ rounds, after which it returns with its reputation reset to the initial value. The suspension duration doubles after each repetition, progressing from $5$ to $10$, $20$, and $40$ rounds. Thus, a node flagged once by chance can rejoin quickly, while a persistently anomalous node is excluded for increasingly longer periods.

\subsection{Phase 5. Leader Election, Aggregation and Block Mining}
\subsubsection{Leader election}
The round leader is elected by a verifiable lottery in which the probability of selection is proportional to reputation,
\begin{equation}
    P_{\mathrm{elect}}(i) = \frac{R_i}{\sum_{n \in \mathcal{N}_{active}} R_n}
\end{equation}
Choosing the highest reputation node deterministically would let one participant monopolise mining and re-centralise the network, whereas a proportional lottery preserves decentralisation while still favouring trustworthy nodes. The lottery uses a Verifiable Random Function \cite{vrf} seeded by the previous block hash, and the winner publishes a proof of selection that any peer can verify.

\subsubsection{Global model aggregation}
Accepted updates are combined by reputation weighted federated averaging \cite{barkatsa2025fair}. Where standard FedAvg weights a contribution by dataset size, PoCQ weights it by demonstrated trustworthiness,
\begin{equation}
    w_g^{t+1} = w_g^t + \sum_{i \,\mid\, \Psi_i=1} \frac{R_i}{\sum_{n \,\mid\, \Psi_n=1} R_n} \, \Delta w_i^t
\end{equation}

\subsubsection{Lightweight block construction}
The leader mines a block whose body excludes model parameters entirely and carries only audit metadata, namely the block index, the commitments and profiles of the round, the vote transactions, the identifiers of the accepted workers, the miner identity, the previous block hash and a timestamp. Keeping parameters out of the consensus record keeps the chain state small and synchronisation fast, while retaining everything a peer needs in order to recompute the magnitude part of any vote. Algorithm~\ref{alg:pocq} summarises the complete round.

\begin{algorithm}[ht!]
		\small
		\caption{PoCQ, Reputation Aware Consensus Mechanism}
		\label{alg:pocq}
		\begin{algorithmic}[1]
			\REQUIRE Nodes $\mathcal{N}$, datasets $\{D_i\}$, initial model $w_g^0$, rounds $T_{\max}$,
			bounds $\tau, \zeta, \nu, \phi_{\min}$, decay $\beta$.
			\ENSURE Final global model $w_g^{T_{\max}}$.
			
			\STATE \textbf{Initialise} $\forall i \in \mathcal{N},\ R_i^0 \leftarrow 0.5$
			
			\FOR{round $t = 1$ \textbf{to} $T_{\max}$}
			\STATE \textcolor{gray}{\textit{// Phase 1. train, commit, then reveal}}
			\FOR{each worker $i \in \mathcal{N}$ \textbf{in parallel}}
			\STATE $\Delta w_i^t \leftarrow \text{SGD}(w_g^{t-1}, D_i) - w_g^{t-1}$
			\STATE $\mathbf{p}_i \leftarrow (\|\Delta w_{i,1}^t\|_2, \dots, \|\Delta w_{i,\Lambda}^t\|_2)$
			\STATE Publish $H_i \leftarrow \text{Hash}(\Delta w_i^t)$ and $\mathbf{p}_i$, signed with $SK_i$
			\ENDFOR
			\STATE Close the commitment block, then flag replays and copies among $\{H_i\}$
			
			\STATE \textcolor{gray}{\textit{// Phase 2. assignment, recomputable from the chain}}
			\STATE $\{\mathcal{V}_i\} \leftarrow \text{Assign}(\mathcal{N}, k_{\min}, k_{\max}, \text{Hash}(B_{t-1}))$
			\STATE Each worker reveals $\Delta w_i^t$ to its validators
			
			\STATE \textcolor{gray}{\textit{// Phase 3. reference based validation}}
			\FOR{each validator $j$, each assigned worker $i$}
			\STATE $\mathcal{R}_{j \to i} \leftarrow \mathcal{H}_j \setminus \{i\}$
			\STATE $\mathbf{m} \leftarrow$ median profile of $\mathcal{R}_{j \to i}$, \ \ $\mathbf{d} \leftarrow$ median direction
			\STATE $r_{\ell} \leftarrow \mathbf{p}_{i,\ell} / \mathbf{m}_{\ell}$ \ for $\ell = 1 \dots \Lambda$
			\STATE $\text{layers\_ok} \leftarrow \sum_{\ell} \mathbb{I}(r_{\ell} > \tau \lor r_{\ell} < \zeta) < \nu$
			\STATE $\phi \leftarrow \cos(\Delta w_i^t, \mathbf{d})$, \ \ $\text{dir\_ok} \leftarrow \phi \ge \phi_{\min}$
			\STATE $z_{j \to i} \leftarrow \mathbb{I}(\text{layers\_ok} \land \text{dir\_ok})$
			\STATE Publish $\mathcal{Z}_{tx}$ carrying $z_{j \to i}$, $\{r_{\ell}\}$ and $\phi$
			\ENDFOR
			
			\STATE \textcolor{gray}{\textit{// Phase 4a. provable violations, settled first}}
			\FOR{each node $n$}
			\IF{commitment or profile mismatch \OR replay \OR copy \OR vote mismatch}
			\STATE $\text{Expel}(n)$, and discard its votes for this round
			\ENDIF
			\ENDFOR
			
			\STATE \textcolor{gray}{\textit{// Phase 4b. consensus and reputation}}
			\FOR{each worker $i$}
			\STATE $S_i \leftarrow \frac{\sum_j z_{j \to i} R_j}{\sum_j R_j}$, \ \ $\Psi_i \leftarrow \mathbb{I}(S_i > 0.5)$
			\IF{$t > T_{\mathrm{warm}}$}
			\STATE $\rho_i \leftarrow +1$ if $\Psi_i = 1$, else $-1$
			\STATE $R_i \leftarrow \max(0, \min(1, \beta R_i + (1-\beta)\rho_i))$
			\ENDIF
			\ENDFOR
			\STATE Suspend nodes with $R_i < R_{\min}$ for $T_{\mathrm{susp}}$ rounds
			
			\STATE \textcolor{gray}{\textit{// Phase 5. leader, aggregation, block}}
			\STATE Elect leader $L$ with $P_{\mathrm{elect}}(L) \propto R_L$ via VRF on $\text{Hash}(B_{t-1})$
			\STATE $w_g^t \leftarrow w_g^{t-1} + \sum_{i \mid \Psi_i=1} \frac{R_i}{\sum_{n \mid \Psi_n=1} R_n} \Delta w_i^t$
			\STATE $L$ mines $B_t$ with commitments, profiles, votes and accepted identifiers
			\ENDFOR
			\RETURN $w_g^{T_{\max}}$
		\end{algorithmic}
\end{algorithm}

\begin{figure}[!t]
\centering
\includegraphics[width=0.5\textwidth]{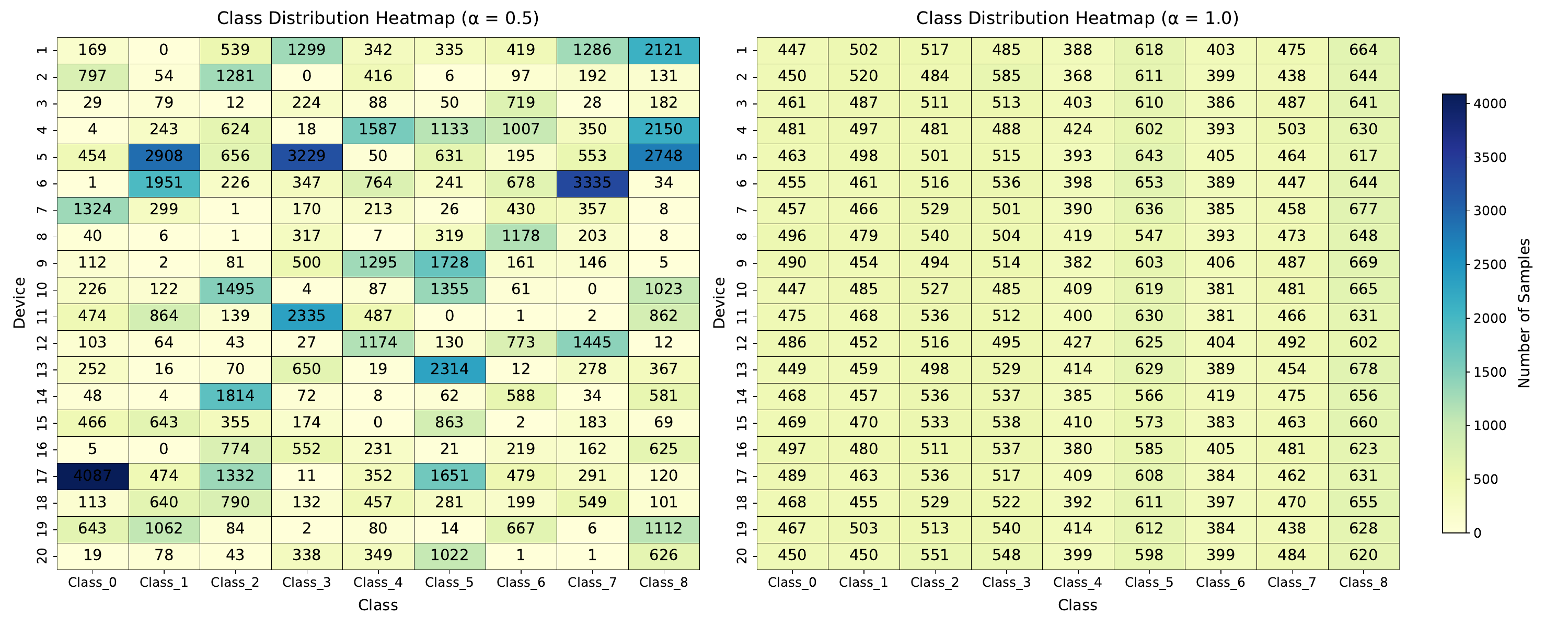}
\caption{Class distribution of PathMNIST across devices at two heterogeneity levels.} 
\label{fig:data_distributions}
\end{figure}

\section{Experiments And Discussion}
\label{sec:experiments}
\subsection{Experiments Setup}
This section describes the dataset and model, the threat model and attacks, the methods compared, the evaluation metrics and the configuration used throughout.

\subsubsection{Dataset and Model Architecture}

All experiments are conducted on PathMNIST, a nine class histological dataset of colon pathology drawn from MedMNIST v2 \cite{Yang2023}. We concentrate the evaluation on a single dataset because the present study broadens the threat model rather than the data, and reports five distinct poisoning attacks, two adversary knowledge models, six competing methods, a sensitivity analysis over attack strength and a stress test over the adversary fraction.

To accurately simulate the inherently unbalanced data distributions present in real-world federated networks, we evaluate the models under both Independent and Identically Distributed (IID) and Non-IID configurations. The variation in data distributions across nodes in the non-IID setting is induced by partitioning the labels across clients using a Dirichlet distribution, $Dir(\alpha)$ \cite{kasyap2023efficient, reguieg2023comparative}.  The non-IID setting uses $\alpha = 0.5$, which represents moderate heterogeneity and is the regime in which a magnitude based rule is genuinely at risk of confusing skew with poisoning, see Fig.~\ref{fig:data_distributions}.

The local model employed on each client is a convolutional neural network (CNN) comprising two convolutional layers with 32 and 64 filters of size 5×5, each followed by ReLU activation and 2×2 max pooling, and two fully connected layers of sizes 512 and the number of classes of each dataset. The model accepts single- or multi-channel input images of size 28×28, making it compatible with all datasets used in our experiments. The network therefore has $\Lambda = 4$ weight matrices, which is the length of the norm profile explained in Section \ref{sec:methodology}, and roughly $1.66$ million parameters.

\subsubsection{Threat Model and Attacks}
\label{sec:threat_model}
All adversaries in this study perform model poisoning by submitting deliberately corrupted parameter updates. Adversarial nodes remain fixed throughout each run, and their number is denoted by $M$. We evaluate five attacks, grouped according to the information available to the adversary.

Two attacks are blind, in the sense that the adversary constructs its submission from its own update alone. Under the Gaussian attack an adversary adds independent noise of scale $\sigma$ to every parameter of its trained update, which inflates the magnitude without changing the direction systematically. Under sign flipping an adversary negates its update, which reverses the direction while leaving every norm exactly unchanged. The two are complementary by construction, and no single scalar statistic detects both.

The remaining three attacks are informed. The adversary receives the full set of benign updates in a round and uses this information to construct a perturbation intended to evade the defence. LIE adds a multiple of the coordinate wise standard deviation to the benign mean, with the multiple selected to keep the result within the range occupied by most benign updates \cite{baruch2019lie}. Min-Max maximises the displacement while ensuring that the crafted update remains no farther from any benign update than benign updates are from one another \cite{shejwalkar2021manipulating}. Fang optimises the perturbation against an assumed aggregation rule by using a surrogate copy of the defence \cite{fang2020local}.

Granting the informed attacks a complete view of the round is a deliberate concession. PoCQ never reveals more than a committee sized subset of the updates to any participant, and the ledger carries only hashes and norm profiles, so the information these attacks consume is not available to a real adversary in this protocol. We supply it anyway so that the defence is measured against the strongest published attacks rather than against the weakened versions its own communication pattern would impose.

Two further behaviours target the consensus layer rather than the model. A bad-mouthing validator votes invalid against every assigned peer, regardless of the observed update. A free-rider submits an update it did not train by replaying its own earlier update or copying a neighbour’s update, either unchanged or with a small perturbation.

\subsubsection{Benchmark Algorithms and Threat Configurations}
\label{Benchmark&ThreatConfig}

We benchmark PoCQ against six methods, which fall into two groups and therefore require different interpretations.

The first group consists of aggregation rules. Vanilla Federated Learning (VFL) applies Federated Averaging without any defence and provides a lower bound for Byzantine resilience \cite{McMahan2017}. Krum selects the single update closest to its nearest neighbours \cite{blanchard2017krum}, while coordinate wise trimmed mean and coordinate wise median aggregate updates independently across each parameter dimension \cite{yin2018byzantine}. As set out in Section \ref{sec:robust_agg}, these methods make no persistent judgement about participants and never exclude them from the network, so their detection metrics are undefined throughout. We evaluate them with a complete view of every update in the round and provide, for Krum and trimmed mean, the true value of $M$. These are the most favourable assumptions available to them, and assumptions the decentralised methods do not receive.

The second group consists of decentralised blockchain frameworks that perform detection and maintain reputation, which is the class PoCQ belongs to. Validation-based Blockchained Federated Learning (VBFL) has validators retrain each received update on their own validation data and vote on the resulting accuracy change \cite{chen2021robust}. Lightweight Blockchained Federated Learning (LBFL) uses a proof-of-contribution committee in which contribution combines a single retraining epoch with the reported dataset size and epoch count \cite{qiao2024lbfl}. These two represent the state of the art in blockchain based decentralised federated learning, and they are selected because each validates individual updates and publishes an operating point, so each can be reproduced at the setting its authors report. They are the direct comparators, and every claim about detection quality in this work is made against them. Frameworks that validate against a shared dataset, such as T-BFL, are not included, because supplying that dataset would change the problem rather than the method. Granting an aggregation rule a full view of the population costs nothing outside the algorithm, whereas granting a shared test set requires the institutions to release labelled records. 

Both frameworks run at the operating point their authors publish, with one exception. Each judges a peer by an accuracy difference measured against a reference the validator trains on its own data, and under a Dirichlet partition that quantity is not centred on zero, since the reference is fitted to the validator's own class mix and the candidate is not. At the published setting this leaves almost no honest update acceptable under $\alpha = 0.5$ once the warm-up ends, and the global model stops advancing. Under non-IID data each framework is therefore given the most permissive setting at which it continues to train, which is a concession in its favour, so its false positive rates there are a lower bound on the cost of these rules under heterogeneity. The published setting is used throughout under IID, where the same quantity is centred on zero.

The simulated network contains 20 nodes throughout. The main experiments consider a clean setting with no adversaries and an adversarial setting with $M = 4$, representing 20\% of the network. Section \ref{sec:stress} increases $M$ to 6, 8, 10, and 11 to identify the point at which each mechanism fails.

\subsubsection{Evaluation Metrics}
\label{sec:metrics}
We evaluate two distinct properties of each defence separately, as strong performance on one does not necessarily imply strong performance on the other.

The first property is whether a poisoned update contributes to the global model. This is measured at the level of an individual update in a single round. A positive outcome denotes an update excluded from the aggregate, either because consensus rejected it or because its owner was excluded from participation at that time. The true positive rate, also called recall, is the proportion of adversarial updates excluded, whereas the false positive rate is the proportion of honest updates excluded. Precision is the fraction of excluded updates that are genuinely adversarial, indicating how accurately the defence excludes malicious updates. These metrics are reported in Table~\ref{tab:detection}.
	
The second property is whether the framework excludes the right participants. This is evaluated at the node level over an entire run. A positive outcome denotes a node blacklisted at any point during the 50 rounds. This property therefore measures whether a node is eventually excluded, not whether its individual updates were rejected, which the first property already covers. These are the quantities in Table~\ref{tab:node_precision}, the confusion matrices of Fig.~\ref{fig:confusion_m} and the timeline of Fig.~\ref{fig:blk_timeline}.

The two must not be conflated. A framework may reject a large share of honest updates every round and still exclude few nodes, and a framework that rejects almost nothing may still remove a node permanently after a long accumulation of small penalties. Accuracy is reported as the mean over the final ten of 50 communication rounds, which removes the round to round variance of a single final measurement.

\subsubsection{Computational Setup and Experiments Configuration}
\label{Comp_setup}
The simulated federated network, cryptographic consensus protocols, and deep learning models were implemented using PyTorch framework. All experiments were executed on an NVIDIA T400 GPU with total of 12 GB of RAM.

We simulate a network of 20 nodes for 50 communication rounds, by which point the global accuracy of all methods has stabilised. Each node performs five local epoch in each round using a batch size of 32.

The validator committee size is bounded by $k_{\min} = 4$ to $k_{\max} = 6$. Section \ref{sec:stress} treats committee size as a  swept parameters. The magnitude test compares each update with the reference median using an upper ratio bound of $\tau = 2.5$ and a lower bound of $\zeta = 0.1$. An update is rejected when at least $\nu = 2$ of the four weight matrices fall outside this band. The direction test uses $\phi_{\min} = -0.1$, which lies below zero so as to avoid penalising updates that are simply uncorrelated with their peers. Reputation is initialised at $0.5$ and updated using a decay factor of $\beta = 0.9$. Reputation remains fixed during the first $T_{\mathrm{warm}} = 5$ rounds. A node whose reputation falls below $R_{\min} = 0.1$ is suspended for $T_{\mathrm{susp}} = 5$ rounds on its first offence, with the duration doubling after each subsequent offence.

The complete source code and simulation configurations are publicly available at \url{https://github.com/sudad/PoCQ.git}.

\subsection{Results and Discussion}
\label{res&dis}

\subsubsection{Accuracy and Robustness}
\label{accuracy}
Table~\ref{tab:accuracy_results} reports the global accuracy of all seven methods across the five attacks and the clean configuration. Three observations follow from it.

Among the decentralised blockchain frameworks, PoCQ has the highest average in both distributions, and the margin is carried by the non-IID panel, where it averages 76.83\% against 76.13\% for VBFL and 70.96\% for LBFL. Both PoCQ and VBFL stand almost six points clear of LBFL. LBFL is last in every non-IID condition including the clean one, which anticipates the pattern examined in the next section, since its accuracy is constrained by the honest updates its own rule rejects rather than by the attack. Under IID the seven methods fall within roughly a point of one another and no method leads more than two of the six rows, so that panel shows that PoCQ costs nothing where a defence is not needed rather than showing an advantage.

The aggregation rules are less consistent. Trimmed mean and coordinate median track the undefended run under the informed attacks but lose about nine points under sign flipping, because four sign reversed updates pull coordinate wise aggregation toward zero in the coordinates where learning occurs. Krum averages 45.97\% under non-IID data, which reflects the rule rather than the strength of the attacks, since it depends on the contribution of only one client and under a Dirichlet partition no individual client holds representative data. Its recovery under IID supports that reading. PoCQ exceeds every aggregation rule on average in both settings without being given a full view of the population.

The informed attacks have a limited effect at this adversary fraction. Undefended Federated Averaging achieves the highest non-IID accuracy under both LIE and Fang, with PoCQ a fraction of a point behind, because these attacks are built to remain within the benign distribution and with four adversaries among 20 participants the corruption they introduce is smaller than the cost of occasionally rejecting an honest update. Their effect becomes more pronounced at higher adversary fractions, as examined in Section~\ref{sec:stress}.

\begin{table}[t]
	\centering
	\caption{Global accuracy (\%) on PathMNIST with 20 nodes and $M=4$, averaged over the final ten of 50 rounds. Best in each
		row in bold.}
	\label{tab:accuracy_results}
	
	\footnotesize
	\setlength{\tabcolsep}{3.5pt}
	\renewcommand{\arraystretch}{0.92}
	
	\begin{tabular}{@{}lccccccc@{}}
		\toprule
		&
		\multicolumn{4}{c}{\textbf{Aggregation rules}}
		&
		\multicolumn{3}{c}{\textbf{Blockchain frameworks}} \\
		\cmidrule(lr){2-5}\cmidrule(lr){6-8}
		\textbf{Attack}
		& VFL & Krum & \shortstack{Trimmed\\Mean} & \shortstack{Coord.\\Median}
		& \textbf{PoCQ} & VBFL & LBFL \\
		\midrule
		
		\multicolumn{8}{c}{\textit{Non-iid, Dirichlet $\alpha=0.5$}} \\
		\midrule
		
		No attack
		& 77.57 & 45.53 & 76.89 & 76.18 & \textbf{77.59} & 76.94 & 70.98 \\
		Fang
		& \textbf{76.76} & 41.88 & 72.90 & 68.84 & 76.37 & 76.33 & 70.66 \\
		Gaussian
		& 74.65 & 44.20 & 75.38 & 74.53 & \textbf{76.81} & 76.68 & 71.15 \\
		LIE
		& \textbf{77.02} & 55.28 & 76.22 & 75.07 & 76.74 & 76.63 & 72.89 \\
		Min-Max
		& 75.55 & 46.43 & 74.64 & 72.21 & \textbf{76.25} & 74.45 & 70.14 \\
		Sign flip
		& 67.94 & 42.51 & 67.65 & 68.16 & \textbf{77.21} & 75.74 & 69.96 \\
		\textbf{Average}
		& 74.91 & 45.97 & 73.95 & 72.50 & \textbf{76.83} & 76.13 & 70.96 \\
		
		\midrule
		
		\multicolumn{8}{c}{\textit{IID}} \\
		\midrule
		
		No attack
		& 78.83 & 69.53 & \textbf{78.93} & 78.46 & 78.81 & 78.46 & 78.39 \\
		Fang
		& 78.40 & 67.73 & 78.96 & 77.49 & \textbf{79.00} & 78.06 & 78.48 \\
		Gaussian
		& 78.28 & 69.86 & 78.62 & 78.31 & \textbf{78.66} & 77.97 & 77.23 \\
		LIE
		& 77.84 & 67.11 & \textbf{78.06} & 76.98 & 77.94 & 77.95 & 75.74 \\
		Min-Max
		& 75.50 & 58.74 & 75.53 & 74.81 & 76.17 & \textbf{77.77} & 77.27 \\
		Sign flip
		& 75.67 & 69.86 & 77.61 & 77.21 & 78.42 & 78.00 & \textbf{78.77} \\
		\textbf{Average}
		& 77.42 & 67.14 & 77.95 & 77.21 & \textbf{78.17} & 78.04 & 77.65 \\
		\bottomrule
	\end{tabular}
\end{table}

\subsubsection{Detection Quality}
\label{blacklisting}
Table~\ref{tab:detection} evaluates whether individual updates reach the global model. The clean configuration establishes the cost of validation in the absence of attack. PoCQ excludes fewer than one percent of honest node rounds in both data settings, whereas VBFL and LBFL exclude between one tenth and four fifths. Because the attacked results remain close to these clean figures, this loss is primarily a property of the validation rule rather than a response to adversarial behaviour.
	
The true positive rate produces a different ranking, with LBFL detecting the largest share of adversarial updates and PoCQ outperforming VBFL. Precision provides the more informative measure because it accounts for the cost of those rejections. PoCQ achieves an average precision of 0.581 under non-IID data and 0.878 under IID data. The best figure reached by VBFL or LBFL in either data setting is 0.391, so PoCQ exceeds it in both. It leads on nine of ten attack conditions. \tanchor{note:lie}The exception is LIE under non-IID data (*), where it rejects no adversarial update at all.
	
This exception reflects the intended design of informed attacks. A crafted update within the benign range also appears benign to a geometric test. The mean cosine similarity of a LIE update to the reference direction is 0.692 against 0.226 for an honest update in the same runs, so such an update is more closely aligned with the consensus direction than honest nodes are. Although LBFL rejects most Min-Max updates, it also excludes most honest node rounds, indicating that its adversarial and honest rejections are not well separated. \tanchor{note:lbfl}The same lack of separation leaves it with a precision of 0.017 under LIE ($\dagger$), where rejecting 0.726 of the honest updates buys it the largest value in that row without producing a usable detection. Table~\ref{tab:accuracy_results} indicates that at $M = 4$ the remaining informed attacks have limited effect on global accuracy.  The blind attacks are the opposite case, and they are why the rule carries two arms. Under Gaussian noise the magnitude test raises 5.28 validator flags per round and the direction test none, and under sign flipping the counts reverse to 0.36 and 6.20, since negation preserves every norm exactly and is visible only as a reversal. \tanchor{note:vbfl}VBFL detects the same Gaussian attack under non-IID data but not under IID data ($\ddagger$). Its vote is the accuracy change a validator measures after retraining the candidate on its own data. Under IID that data is representative, so a single epoch absorbs noise that pushes the model in no particular direction. No accuracy change then remains to measure, and the adversarial updates score below the honest ones. Under non-IID the narrow local data cannot repair the perturbed model, so the change stays visible. A norm profile cannot be undone by retraining in this way, which is why PoCQ detects the attack in both settings.
	
\begin{figure}[ht]
		\centering
		\includegraphics[width=\linewidth]{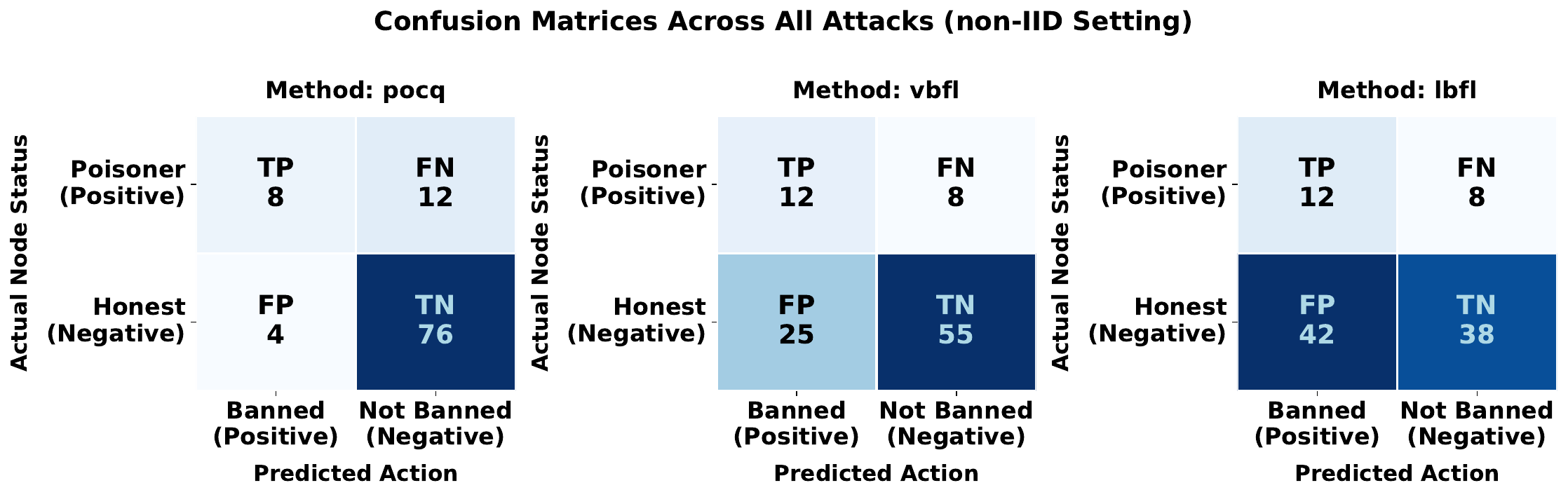}
		\caption{Node level confusion matrices for the runs summarised in Table~\ref{tab:node_precision}.}
		\label{fig:confusion_m}
	\end{figure}

\begin{figure*}[ht]
		\centering
		\includegraphics[width=0.85\linewidth]{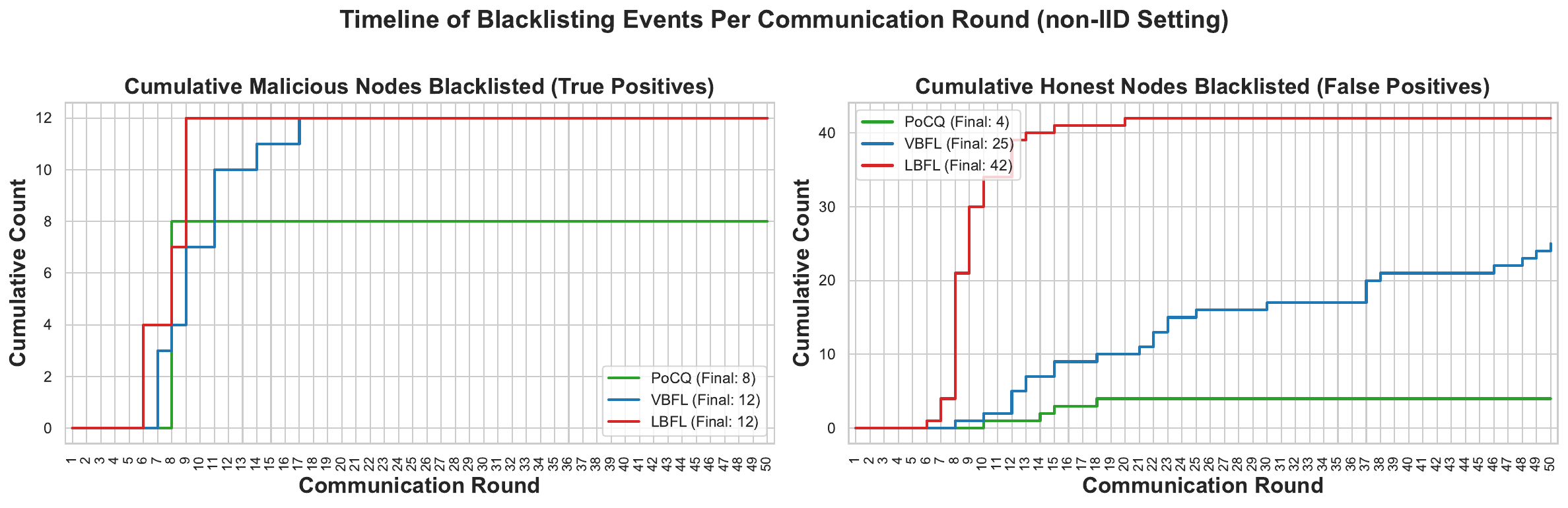}
		\caption{Cumulative blacklisting per round under non-IID data. Left, adversaries removed. Right, honest nodes removed.}
		\label{fig:blk_timeline}
	\end{figure*}

Figure~\ref{fig:confusion_m} and Table~\ref{tab:node_precision} address the second property of Section~\ref{sec:metrics}, which participants are blacklisted. VBFL and LBFL exclude more adversaries, but they also exclude 25 and 42 honest node instances against 4 for PoCQ. PoCQ is therefore the only method whose exclusions are right more often than wrong, at a precision of 0.667 against 0.324 and 0.222, an improvement of 106\% over the stronger of the two. Its recall is two thirds of theirs, but the two errors do not cost the same. An adversary that is never suspended is not thereby accepted, since its updates face the test in every round, and Table~\ref{tab:detection} shows that PoCQ excludes above nine tenths of them under the blind attacks. An excluded hospital leaves the federation with its data. Every PoCQ exclusion is in any case temporary, the last occurring by round 18 (Fig.~\ref{fig:blk_timeline}), since statistical anomalies draw a bounded suspension and expulsion is reserved for violations provable from the ledger. PoCQ further reaches the highest non-IID average of the three frameworks in Table~\ref{tab:accuracy_results} while excluding the fewest participants.

	\begin{table}[t]
		\centering
		\caption{Node level precision and recall under non-IID data, pooled over the five attacks.
			A positive is a node excluded at any point in the run.}
		\label{tab:node_precision}
		\footnotesize
		\setlength{\tabcolsep}{5pt}
		\renewcommand{\arraystretch}{0.95}
		\begin{tabular}{@{}lcccc@{}}
			\toprule
			\textbf{Method} & \shortstack{\textbf{Adversaries}\\\textbf{excluded}}
			& \shortstack{\textbf{Honest}\\\textbf{excluded}}
			& \textbf{Precision} & \textbf{Recall} \\
			\midrule
			\textbf{PoCQ} & 8 of 20  & 4 of 80  & \textbf{0.667} & 0.400 \\
			VBFL          & 12 of 20 & 25 of 80 & 0.324 & 0.600 \\
			LBFL          & 12 of 20 & 42 of 80 & 0.222 & 0.600 \\
			\bottomrule
		\end{tabular}
	\end{table}

	\begin{table}[t]
		\centering
		\caption{Update level detection quality with 20 nodes and $M=4$. Best precision in each row in bold. Marked values are discussed in the text.}
		\label{tab:detection}
		\scriptsize
		\renewcommand{\tmarkscale}{1.2}
		\setlength{\tabcolsep}{2.0pt}
		\renewcommand{\arraystretch}{0.90}
		
		\begin{tabular}{@{}l*{9}{c}@{}}
			\toprule
			&
			\multicolumn{3}{c}{\textbf{TPR}}
			&
			\multicolumn{3}{c}{\textbf{FPR}}
			&
			\multicolumn{3}{c}{\textbf{Precision}} \\
			\cmidrule(lr){2-4}\cmidrule(lr){5-7}\cmidrule(lr){8-10}
			\textbf{Attack}
			& PoCQ & VBFL & LBFL & PoCQ & VBFL & LBFL  & PoCQ & VBFL & LBFL \\
			\midrule
			
			\multicolumn{10}{c}{\textit{Non-iid, Dirichlet $\alpha=0.5$}} \\
			\midrule
			No attack & -- & -- & -- & 0.005 & 0.104 & 0.808 & -- & -- & -- \\
			\cmidrule(lr){1-10}
			Fang
			& 0.075 & 0.000 & 0.000 & 0.046 & 0.101 & 0.758 & \textbf{0.290} & 0.000 & 0.000 \\
			Gaussian & 0.930 & 0.511 & 0.918 & 0.003 & 0.049 & 0.773 & \textbf{0.989} & 0.763 & 0.294 \\
			LIE & 0.000 & 0.000 & 0.077 & 0.001 & 0.078 & 0.726 & 0.000\tmark{note:lie}{*} & 0.000 & \textbf{0.017}\tmark{note:lbfl}{$\dagger$} \\
			Min-Max & 0.050 & 0.250 & 0.964 & 0.007 & 0.078 & 0.778 & \textbf{0.625} & 0.465 & 0.306 \\
			Sign flip & 0.910 & 0.686 & 0.923 & 0.000 & 0.078 & 0.791 & \textbf{1.000} & 0.727 & 0.289 \\
			\textbf{Avg. (attacks)} & 0.393 & 0.289 & 0.576 & 0.011 & 0.077 & 0.765 & \textbf{0.581} & 0.391 & 0.181 \\
			
			\midrule
			
			\multicolumn{10}{c}{\textit{IID}} \\
			\midrule
			No attack & -- & -- & -- & 0.000 & 0.400 & 0.504 & -- & -- & -- \\
			\cmidrule(lr){1-10}
			Fang & 0.200 & 0.214 & 0.913 & 0.003 & 0.425 & 0.510 & \textbf{0.952} & 0.104 & 0.386 \\
			Gaussian & 1.000 & 0.000\tmark{note:vbfl}{$\ddagger$} & 0.308 & 0.001 & 0.333 & 0.402 & \textbf{0.995} & 0.000 & 0.153 \\
			LIE & 0.020 & 0.000 & 0.000 & 0.006 & 0.320 & 0.559 & \textbf{0.444} & 0.000 & 0.000 \\
			Min-Max & 0.060 & 0.832 & 0.913 & 0.000 & 0.353 & 0.407 & \textbf{1.000} & 0.392 & 0.441 \\
			Sign flip & 0.765 & 0.671 & 0.944 & 0.000 & 0.355 & 0.402 & \textbf{1.000} & 0.361 & 0.457 \\
			\textbf{Avg. (attacks)} & 0.409 & 0.343 & 0.616 & 0.002 & 0.357 & 0.456 & \textbf{0.878} & 0.171 & 0.287 \\
						\bottomrule
		\end{tabular}
	\end{table}
	
	\begin{table}[ht]
		\centering
		\caption{Effect of the Gaussian noise scale $\sigma$ under non-IID data with $M=4$. The two PoCQ detection columns are update level, defined as in Table~\ref{tab:detection}. A dash marks a cell for which the corresponding run was not executed.}
		\label{tab:sigma}
		\small
		\setlength{\tabcolsep}{4pt}
		\resizebox{\columnwidth}{!}{%
			\begin{tabular}{l ccc cccc cc}
				\toprule
				& \multicolumn{3}{c}{\textbf{PoCQ}} & \multicolumn{4}{c}{\textbf{Aggregation rules}} & \multicolumn{2}{c}{\textbf{Blockchain}} \\
				\cmidrule(lr){2-4}\cmidrule(lr){5-8}\cmidrule(lr){9-10}
				$\sigma$ & Acc. & TPR & FPR & VFL & Krum & Trim. & Med. & VBFL & LBFL \\
				\midrule
				0.0005 & 74.65 & 0.020 & 0.003 & -- & -- & -- & -- & -- & -- \\
				0.001  & 74.61 & 0.050 & 0.004 & -- & -- & -- & -- & -- & -- \\
				0.002  & 76.26 & 0.130 & 0.001 & 74.33 & 44.20 & 74.46 & 73.54 & 76.94 & 70.17 \\
				0.004  & 76.81 & 0.930 & 0.003 & 74.65 & 44.20 & 75.38 & 74.53 & 76.68 & 71.15 \\
				0.008  & 77.15 & 1.000 & 0.003 & -- & -- & -- & -- & -- & -- \\
				\bottomrule
			\end{tabular}%
		}
	\end{table}
	
\subsubsection{Sensitivity to Attack Strength}
A geometric test has an operating range, and Table~\ref{tab:sigma} locates it by sweeping the Gaussian noise scale over a sixteen fold range. Detection is close to a step function. At $\sigma \le 0.002$ PoCQ keeps out between 2\% and 13\% of the adversarial updates and blacklists no adversary at all. At $\sigma \ge 0.004$ it keeps out 93\% and then 100\% of them and blacklists all four. The false positive rate is between 0.001 and 0.004 at every point in the sweep, so the transition costs nothing in honest traffic.

The interesting property is what happens to accuracy across that transition. PoCQ scores 74.65\% at $\sigma = 0.0005$, where it detects nothing, and 77.15\% at $\sigma = 0.008$, where it detects everything. Accuracy rises with the strength of the attack, because an attack weak enough to evade a magnitude test is also weak enough to be absorbed by reputation weighted averaging over sixteen honest updates. The undefended run confirms this reading, since VFL scores 74.33\% at $\sigma = 0.002$ and 74.65\% at $\sigma = 0.004$, so the noise at these scales is not what limits accuracy. The region below the detection threshold is therefore not a region in which PoCQ is vulnerable. It is a region in which there is little to defend against.

The contrast at the same operating point is instructive. At $\sigma = 0.002$ VBFL blacklists all four adversaries and four honest nodes, and LBFL blacklists two adversaries and nine honest nodes, whereas PoCQ blacklists nobody. The resulting accuracies are 76.94\%, 70.17\% and 76.26\%. VBFL converts its greater sensitivity into a 0.68 point advantage on this one row, and Fig.~\ref{fig:confusion_m} shows what the same sensitivity costs it across the rest of the evaluation.

\begin{figure}[ht]
    \centering
    \includegraphics[width=\linewidth]{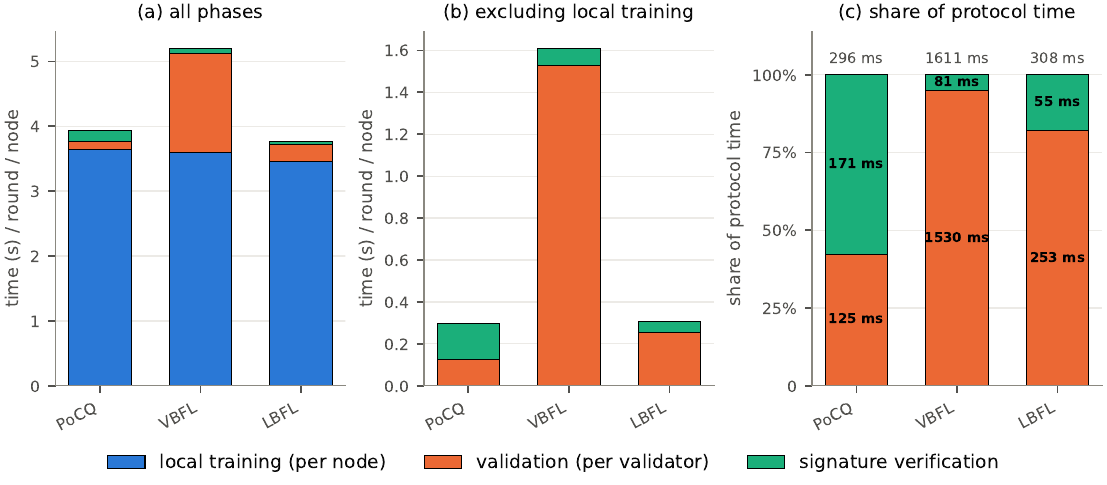}
    \caption{Mean wall clock time per node per round.(a) includes local training, (b) excludes it, (c) gives the composition of protocol time.}
    \label{fig:val_time}
\end{figure}

\subsubsection{Computational Overhead and Validation Efficiency}
\label{validation}
Fig.~\ref{fig:val_time} reports wall clock time measured per node per round on identical hardware, which is the quantity a participant experiences, rather than a sum over the network that grows with the number of participants.

Validation costs 125 ms per validator per round in PoCQ, against 253 ms in LBFL and 1530 ms in VBFL. The ratio follows directly from what the three rules compute. A PoCQ validator forms four layer norms and one cosine over the parameter vector, which is a small number of linear passes and no gradient computation. LBFL retrains each received update for one epoch on a reduced local set. VBFL retrains and then evaluates on a full local validation set, which is why it is an order of magnitude above both, and why the gap would widen with model size rather than narrow.

Counting signature verification as well, the complete protocol outside local training costs 296 ms per node per round in PoCQ, 308 ms in LBFL and 1611 ms in VBFL. The margin over LBFL is therefore small in total even though the validation step is twice as fast, because PoCQ verifies more signatures, 171 ms against 55 ms. This is the price of the commit and reveal structure, which requires a signature on the commitment as well as on the revealed update and on every vote, and it is the mechanism by which free-riding and bad-mouthing become provable rather than inferred. Fig.~\ref{fig:val_time}(a) places all of this in context, since local training accounts for roughly 3.6 s per node per round in every framework, so the entire PoCQ protocol adds about 8\% to the cost of the round, against 45\% for VBFL.

\begin{figure}[ht]
	\centering
	\includegraphics[width=\linewidth]{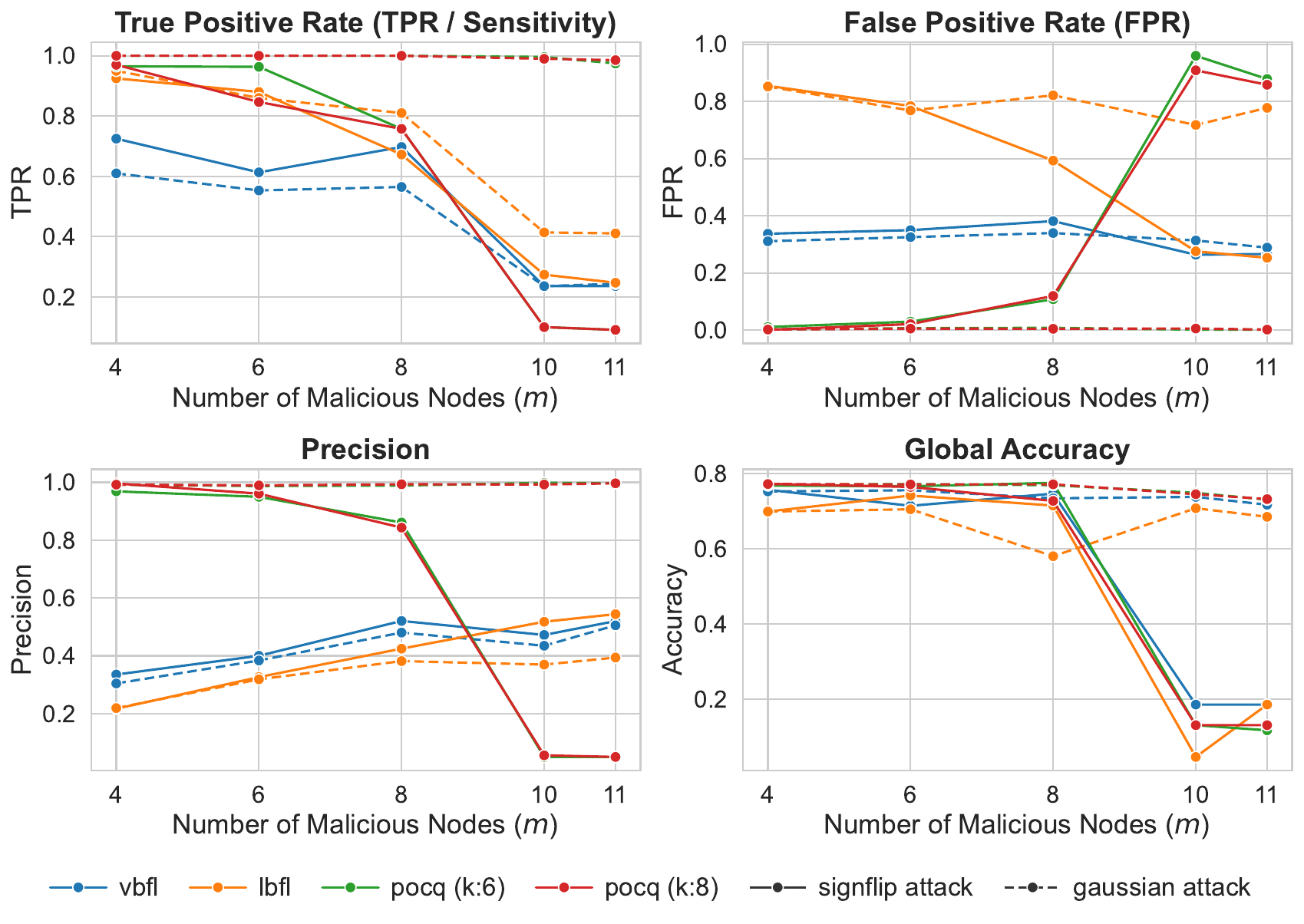}
	\caption{Detection quality and Accuracy as the Malicious count increases. PoCQ at two committee sizes.}
	\label{fig:stress_detect}
\end{figure}

\subsubsection{Robustness Boundary under Stress Test}
\label{sec:stress}
A defence should be assessed not only by its performance within its operating range, but also by the point at which that range fails. We therefore increase the number of adversaries from 4 to 11 in a 20 node network under the two blind attacks, while testing two committee sizes. Figure~\ref{fig:stress_detect} reports the resulting detection behaviour to reports detection quality and global accuracy.

PoCQ remains robust to Gaussian noise throughout the sweep. Almost every poisoned update is rejected throughout the sweep, fewer than one honest update in a hundred is rejected, and accuracy falls by four points, from 77.2\% to 73.1\%. This robustness follows from the independence of the noise. Gaussian attackers do not share a common alternative direction, so they cannot shift the median reference as a coalition. Their inflated norms remain individually visible, while peers rejected in one round are excluded from the reference set in the next. Consequently, even when adversaries form a majority of the network, they rarely form a majority within a validator’s active reference set.

Sign flipping produces a different result. PoCQ remains effective up to $M = 8$ where it holds accuracy of 77.6\% with FPR 0.11, but fails at $M = 10$ with accuracy at 13.2\% with FPR 0.96, and VBFL and LBFL fall to 18.6\% and 4.7\% in the same step. Unlike independent noise, sign flipping gives all adversaries the same reversed direction. Once they constitute a majority within a reference set, they shift the median towards this alternative consensus and cause honest updates to appear anomalous. The resulting false positives rapidly exclude most honest participants. Enlarging the committee does not resolve this problem because the reference is drawn from the active population. A larger committee therefore does not reduce the influence of adversaries. It only reflects their large share of the network more clearly.

The operating boundary is consequently $M < n/2$. Beyond this point, none of the evaluated methods remains robust, including the existing decentralised frameworks and the aggregation rules that receive a full population view. This is not a tuning failure but a property of median based validation, since a median is defined by the majority of the sample it is computed over. Robustness beyond this threshold would require an external trusted reference, such as the clean root dataset used by FLTrust \cite{cao2021fltrust}, which reintroduces the trusted component that decentralisation seeks to avoid.

\subsubsection{Bad-mouthing and Free-riding}
\label{sec:consensus_attacks}
The two attacks directed at the consensus layer are resolved through recomputation rather than statistical inference. Table~\ref{tab:consensus_attacks} shows that bad-mouthing validators are reliably identified because the commitment profiles and committee assignments are public. Any node can recompute the magnitude component of a validator’s vote and compare it with the published vote. All bad-mouthing validators were removed across the three coalition sizes examined, with no honest removals. Accuracy also remained close to the clean reference when malicious validators represented 45\% of the network. This result is independent of the number of malicious validators because each violation is established by the validator’s own signed publication and not the opinion of its peers.

\begin{table}[ht]
		\centering
		\caption{Consensus layer attacks on PoCQ under non-IID data. Accuracy is the global model accuracy of Table~\ref{tab:accuracy_results}. The combined row holds six bad-mouthing validators and four Fang poisoners, and Removed counts the bad-mouthers only. Marked values are discussed in the text.}
		\label{tab:consensus_attacks}
		\small
		\setlength{\tabcolsep}{4pt}
		\resizebox{\columnwidth}{!}{%
			\begin{tabular}{l c c l c c}
				\toprule
				\textbf{Scenario} & \textbf{Adv.} & \textbf{Removed} & \textbf{Evidence} & \textbf{Honest lost} & \textbf{Global acc. (\%)} \\
				\midrule
				Bad-mouthing & 3 & 3 & vote mismatch, provable & 0 & 76.65 \\
				Bad-mouthing & 6 & 6 & vote mismatch, provable & 0 & 76.77 \\
				Bad-mouthing & 9 & 9 & vote mismatch, provable & 0 & 75.80 \\
				Bad-mouthing with Fang\tmark{note:bmfang}{$\mathsection$} & 6 + 4 & 6 & vote mismatch, provable & 4 & 67.72 \\
				\midrule
				Free-riding, exact replay & 4 & 4 & replay, provable & 0 & 77.15 \\
				Free-riding, exact copy & 4 & 4 & copy, provable & 0 & 77.65 \\
				Free-riding, perturbed replay & 4 & 2 & reputation, statistical & 0 & 74.93 \\
				Free-riding, perturbed copy & 4 & 0 & none & 0 & 77.20 \\
				\bottomrule
			\end{tabular}%
		}
	\end{table}

\tanchor{note:bmfang}The combination of bad-mouthing and Fang attack represents the least favourable setting ($\mathsection$). All six bad-mouthing validators are removed by recomputation, but the four Fang poisoners remain, since an informed attack inside the benign range is what a statistical test cannot separate. Removals therefore reduce the active participants, and with four honest nodes suspended, the adversaries hold 4 of the 10 active nodes rather than 4 of 20. Accuracy falls to 67.72\%, reaching the boundary of Section \ref{sec:stress}.
	
Free-riding distinguishes violations that the ledger can resolve directly from those requiring statistical evidence. Exact replays and copied updates produce commitments already present on the ledger, and all such cases are removed by direct lookup without affecting accuracy. A small perturbation creates a new commitment and avoids this check. The lower magnitude bound $\zeta$ detects some of these cases because the resulting update is too small to reflect genuine training, but a perturbed copy of a neighbour’s current update remains undetected. This is a limitation of the mechanism. However, the undetected attack produces no measurable accuracy loss, at 77.20\% against the clean reference of 77.59\%, because it remains close to the copied honest update.

\subsubsection{Limitations}
The results reveal three limitations that should be stated explicitly.

First, Section~\ref{sec:stress} identifies a fundamental robustness boundary. Validation against a reference drawn from the participant population fails once malicious nodes hold a majority of that population, which occurs when the adversary controls half of the network. All evaluated methods share this limitation, including PoCQ.

The second is the low sensitivity to informed attacks reported in Table~\ref{tab:detection}. A geometric test cannot identify an update deliberately designed to remain within the benign range. At four adversaries among 20 participants, the resulting impact is smaller than the cost of rejecting honest updates. However, this margin narrows as the adversary fraction grows.

The third is the undetected perturbed copy of Section~\ref{sec:consensus_attacks}. Commitment binding identifies exact replays and copies, and the lower magnitude bound detects updates too small to reflect genuine training. However, an adversary can alter a copied update enough to change its hash while preserving its norm, and such an update avoids both checks.

\section{Conclusion and Future Work}
\label{sec:conclusion_future_work}
This paper presented PoCQ, a framework for trustworthy decentralised federated learning in which updates are committed before disclosure and judged by small assigned committees against the updates each committee already holds. Experiments on PathMNIST with 20 nodes and five poisoning attacks show that PoCQ achieves the strongest overall accuracy among the decentralised blockchain frameworks while excluding substantially fewer honest updates and participants than VBFL and LBFL. Its geometric validation requires 125 ms per validator per round, and the complete protocol adds approximately 8\% to the cost of a training round. Consensus-layer attacks are also addressed through recomputation of published evidence. Even when nine of 20 validators cast dishonest votes, PoCQ removes the malicious validators without excluding honest nodes.

Three limitations point to the work that follows. Validation against a reference drawn from the participants fails once malicious nodes hold a majority, so a higher boundary requires a reference that a majority cannot move, such as evidence accumulated across rounds rather than judged within a single one. Informed attacks placed inside the benign range call for a signal beyond geometry, such as the consistency of a node's own trajectory over time. A perturbed copy requires binding an update to the work that produced it rather than to its final norm.

\bibliographystyle{elsarticle-num} 
\bibliography{bibliography}

\end{document}